\documentclass[onecolumn,nofootinbib,a4paper,accepted=2021-01-25]{quantumarticle}
\pdfoutput=1
\usepackage[utf8]{inputenc}
\usepackage[T1]{fontenc}
\usepackage{graphics}
\usepackage{url}
\usepackage{hyperref}
\usepackage{xcolor}
\usepackage{bm}
\usepackage{physics}
\usepackage{mathtools}
\usepackage{amssymb}
\usepackage{amsmath}
\newcommand{\eq}[2]{\begin{equation}\begin{split} \label{eq:#1} #2 \end{split}\end{equation}}

\DeclareMathOperator*{\argmin}{arg\,min}
\DeclareMathOperator{\arctantwo}{arctan2}
\usepackage{qcircuit}
\newcommand{\myCirc}[1]{\mbox{\Qcircuit @C=1.em @R=1.4em {#1}}}

\usepackage{algorithm}
\usepackage[noend]{algpseudocode}

\usepackage[symbol]{footmisc}

\usepackage[backend=bibtex, natbib=true, style=numeric-comp, sorting=none, isbn=false, url=false]{biblatex}
\begin{document}
  
  \title{Structure optimization for parameterized quantum circuits}
  
  \author{Mateusz Ostaszewski}
  \email{mm.ostaszewski@gmail.com}
  \orcid{0000-0001-7915-6662}
  \affiliation{Institute of Theoretical and Applied Informatics, Polish Academy of Sciences, Ba{\l}tycka  5, 44-100 Gliwice, Poland}
  \affiliation{Department of Computer Science, University College London, WC1E 6BT London, United Kingdom}
  
  \author{Edward Grant}
  \email{edward.grant@rahko.ai}
  \orcid{0000-0003-0657-1915}
  \affiliation{Department of Computer Science, University College London, WC1E 6BT London, United Kingdom}
  \affiliation{Rahko Limited, N4 3JP London, United Kingdom}
  
  \author{Marcello Benedetti}
  \email{marcello.benedetti@cambridgequantum.com}
  \orcid{0000-0003-0231-1729}
  \affiliation{Department of Computer Science, University College London, WC1E 6BT London, United Kingdom}
  \affiliation{Cambridge Quantum Computing Limited, CB2 1UB Cambridge, United Kingdom}

  \begin{abstract}
    We propose an efficient method for simultaneously optimizing both the structure and parameter values of quantum circuits with only a small computational overhead. Shallow circuits that use structure optimization perform significantly better than circuits that use parameter updates alone, making this method particularly suitable for noisy intermediate-scale quantum computers. We demonstrate the method for optimizing a variational quantum eigensolver for finding the ground states of Lithium Hydride and the Heisenberg model in simulation, and for finding the ground state of Hydrogen gas on the IBM Melbourne quantum computer. 
  \end{abstract}
  
  \maketitle
  
  \section{Introduction}
  
  Methods for tuning the parameters of a quantum circuit to perform a specific task have several important applications. For example, these have been demonstrated in chemical simulation, combinatorial optimization, generative modeling and \mbox{classification~\cite{peruzzo2014variational,farhi2017quantum,farhi2018classification,benedetti2019generative,benedetti2019adversarial,chen2018universal,grant2018hierarchical,benedetti2019parameterized}.} These tasks are usually performed by selecting a fixed circuit structure, parameterizing it using rotation gates, then iteratively updating the parameters to minimize an objective function estimated from measurements. Circuits of this type are known as parameterized quantum circuits. These methods are particularly promising for use with noisy intermediate-scale quantum (NISQ) computers because of their relative tolerance to noise compared to many other quantum algorithms. 
  
  In recent years a lot of progress has been made in improving the performance of parameterized quantum circuits including methods for calculating parameter gradients, hardware efficient ans{\"a}tze, reducing the number of measurements required, and resolving problems with \mbox{vanishing gradients~\cite{mitarai2018quantum,kandala2017hardware,mitarai2019generalization,izmaylov2019revising,grant2019initialization}}. Despite progress, many challenges remain. One of the most critical challenges is addressing the effects of noise. Generally, the effects of noise increase with the depth of the quantum circuit. For this reason it is highly desirable for a parameterized quantum circuit to be as shallow as possible whilst being expressive enough to perform the task at hand.
  
  Few approaches have been proposed for optimizing the structure of a quantum circuit. In Ref.~\cite{li2017approximate} the authors propose using a genetic algorithm for optimizing the circuit structure by selecting candidate gates from a set of allowed gates that are not parameterized. In Ref.~\cite{grimsley2018adapt} the authors propose growing the circuit by iteratively adding parameterized gates and re-optimizing the circuit using gradient descent. 
  
  In this work we propose a method for efficiently optimizing the structure of quantum circuits while optimizing an objective function, for example the energy given by a Hamiltonian operator. Gates are defined so that both the direction and angle of rotation are degrees of freedom. Whilst the angle is parameterized in a continuous manner, the direction is chosen from a set of generators, namely tensor products of Pauli matrices. Optimization is performed by selecting the first gate and finding the direction and angle of rotation that yield minimum energy. This is performed iteratively for all the parameterized gates in the circuit. Once the last gate has been optimized the cycle is repeated until convergence. 
  
  In the Methods section we describe the approach in generality and then provide algorithms for the case of parameterized single-qubit gates and fixed two-qubit gates. This special case is interesting because it can be executed on all existing NISQ hardware and can provide a significant advantage as we show in the Results section. For single-qubit rotations about a fixed direction, the optimal angle can be found using three energy estimations. Independently from our work this property has been used as part of proposed methods for optimizing angles of rotation in quantum circuits~\cite{nakanishi2019sequential,parrish2019jacobi}\footnote[3]{To the best of our knowledge this idea appeared in Nakanishi et al. (March 2019), Parrish et al. (April 2019) and this work (May 2019). Our manuscript was originally titled `Quantum circuit structure learning'.}. Our method extends this and finds also the optimal direction within a set of canonical ones. This comes with very little overhead, since only seven energy estimations are required. Although we demonstrate the method on parameterized single-qubit gates and fixed entangling gates, the method can be applied to higher order gates. In the Discussion section we present possible extensions and ways to reduce the number or energy estimations required for our algorithms. We leave all the details to the Appendix.

  \section{Methods}
  
  Let us consider an optimization problem where the objective function is encoded in a Hermitian operator $M$ and the candidate solution is encoded in a parameterized quantum circuit $U = U_D \cdots U_1$ acting on an $n$-qubit initial state $\rho$. Each gate is either fixed, e.g., a controlled-Z, or parameterized of the form ${ U_d = \exp(-i\tfrac{\theta_d}{2} H_d) }$. Here, $\theta_d \in (-\pi,\pi]$ are angles of rotation and $H_d$ are Hermitian and unitary matrices, e.g., tensor products of Pauli matrices. We collect these parameters into a real vector $\bm{\theta}$ and a vector of matrices $\bm{H}$, respectively.
  
  Without loss of generality we consider the task of minimizing the objective function which we simply refer to as \textit{the energy}. Using subscripts to indicate arguments of functions, we can state the problem as $(\bm{\theta}^*, \bm{H}^*) = \argmin_{\bm{\theta},\bm{H}} \expval{M}_{\bm{\theta},\bm{H}}$, where $\expval{M}_{\bm{\theta}, \bm{H}} = \tr \left (M U \rho U^\dag \right)$.
  
  To solve the problem we present two methods. The first fixes the circuit structure and optimizes only the angles; the second optimizes structure and angles simultaneously. Both methods rely on the fact that the expectation value as a function of an angle of rotation has sinusoidal form. That is, if we fix $\bm{H}$ as well as all the degrees of freedom in vector $\bm{\theta}$ except one, say $\theta_d$, we can express the expectation as $\expval{M}_{\theta_d} = A \sin(\theta_d + B) + C$ where $A,B$ and $C$ are unknown coefficients. A detailed derivation is provided in Appendix~\ref{s:sinusoidal}. Clearly, if we were able to estimate these coefficients, we could also characterize the sinusoidal form. This can be exploited to design opimization strategies.
  
  Our first method is a coordinate optimization~\cite{bertsekas1999nonlinear,saha2010finite,wright2015coordinate} algorithm applied to the angles of rotation. It finds the optimal angle for one gate while fixing all others to their current values, and sequentially cycles through all gates. This is rather simple to perform since at each step the energy has sinusoidal form with period $2\pi$. For gate $U_d$, the optimal angle has a closed form expression
  \eq{coord_minimum}{
    \theta_d^* &= \argmin_{\theta_d} \expval{M}_{\theta_d} \\
    &= \phi - \tfrac{\pi}{2} - \arctantwo \big( 2\expval{M}_{\phi} - \expval{M}_{\phi+\frac{\pi}{2}} - \expval{M}_{\phi-\frac{\pi}{2}} , ~ \expval{M}_{\phi+\frac{\pi}{2}} - \expval{M}_{\phi-\frac{\pi}{2}} \big) + 2\pi k,
  }
  for any real $\phi$ and any integer $k$. In practice we select $k$ such that $\theta_d^* \in (-\pi, \pi]$.
  
  The optimal angle can be found for all $d = 1,\dots,D$ in order to complete a cycle. Once all angles have been updated, a new cycle is initialized unless a stopping criterion is met. A number of potential stopping criteria could be used here. For example, one could stop after a fixed budget of $K_1$ cycles with the caveat that there may still be room for improvement. As another example, one could stop when the energy has not been significantly lowered for $K_2$ consecutive cycles. We call this algorithm \texttt{Rotosolve} and summarize it in Algorithm~\ref{alg:rotosolve}.
  
  The choice of circuit structure is often based on prior knowledge about the problem as well as hardware constraints, e.g., qubit-to-qubit connectivity and gate set. It is reasonable to expect the chosen circuit structure to be suboptimal in the majority of cases. We believe there is large room for improvement here. Our second method relaxes the constraint of the fixed circuit structure and optimizes it along with the angles. Similarly to the first method, we opt for a greedy approach optimizing one gate at a time.
  
  Recall that in the parameterization considered here, $n$-qubit gates are generated by Hermitian and unitary matrices such as tensor products of Pauli matrices $H_d \in \{I,X,Y,Z\}^{\otimes n}$. Structure optimization aims at finding the optimal set of generators, which is clearly a daunting combinatorial problem. The general approach consists of using the expression in Eq.~\eqref{eq:coord_minimum} to find minimizers $\theta^*_d(P) = \argmin_{\theta_d} \expval{M}_{\theta_d , P}$ for all generators $P \in \{I,X,Y,Z\}^{\otimes n}$. Here the second subscript indicates that $P$ is used as the generator for the $d$-th gate. Then, $H_d$ is set to the generator giving lowest energy, and $\theta_d$ is set to the corresponding minimizer. This is repeated for $d=1,\dots,D$ to complete a cycle and the algorithm iterates until a stop criterion is triggered.
  
  It is worth noting that to select the generator we need to know the energy attained by the corresponding optimal angle. In other words, given a generator $P$ and its optimal angle $\theta_d^*(P)$, we need to calculate $\expval{M}_{\theta_{d}^*(P),P} = -A + C$. This can be extrapolated at no additional cost using the expressions for $A$ and $C$ provided in Appendix~\ref{s:sinusoidal}.
  
\begin{figure}
    \begin{minipage}[t]{0.475\linewidth}
      \centering
      \includegraphics[width=.82\linewidth]{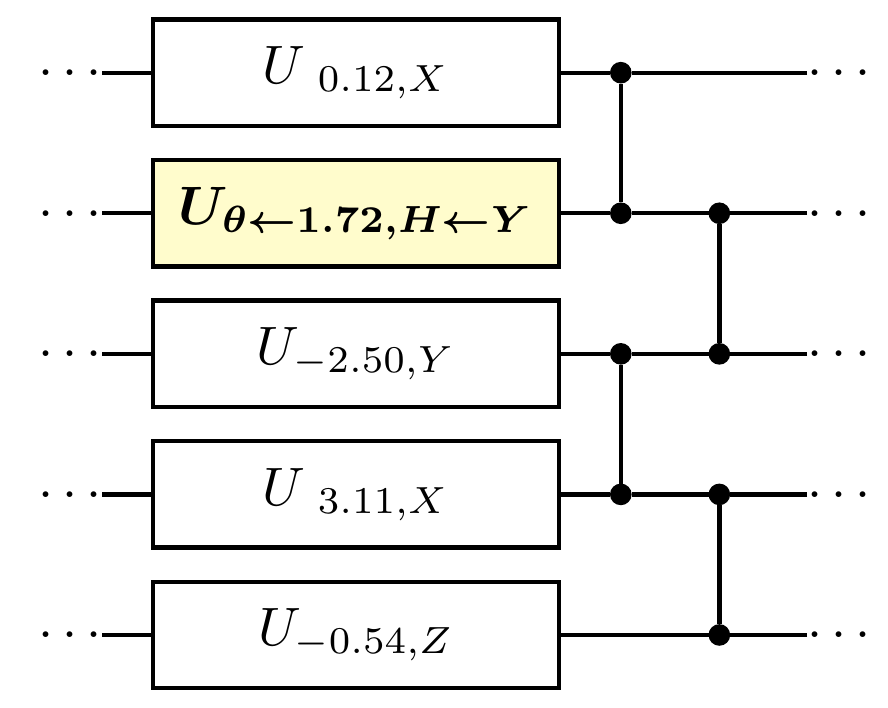}
    \end{minipage}\hfill
    \begin{minipage}[t]{0.475\linewidth}
      \centering 
      \includegraphics[width=.98\linewidth]{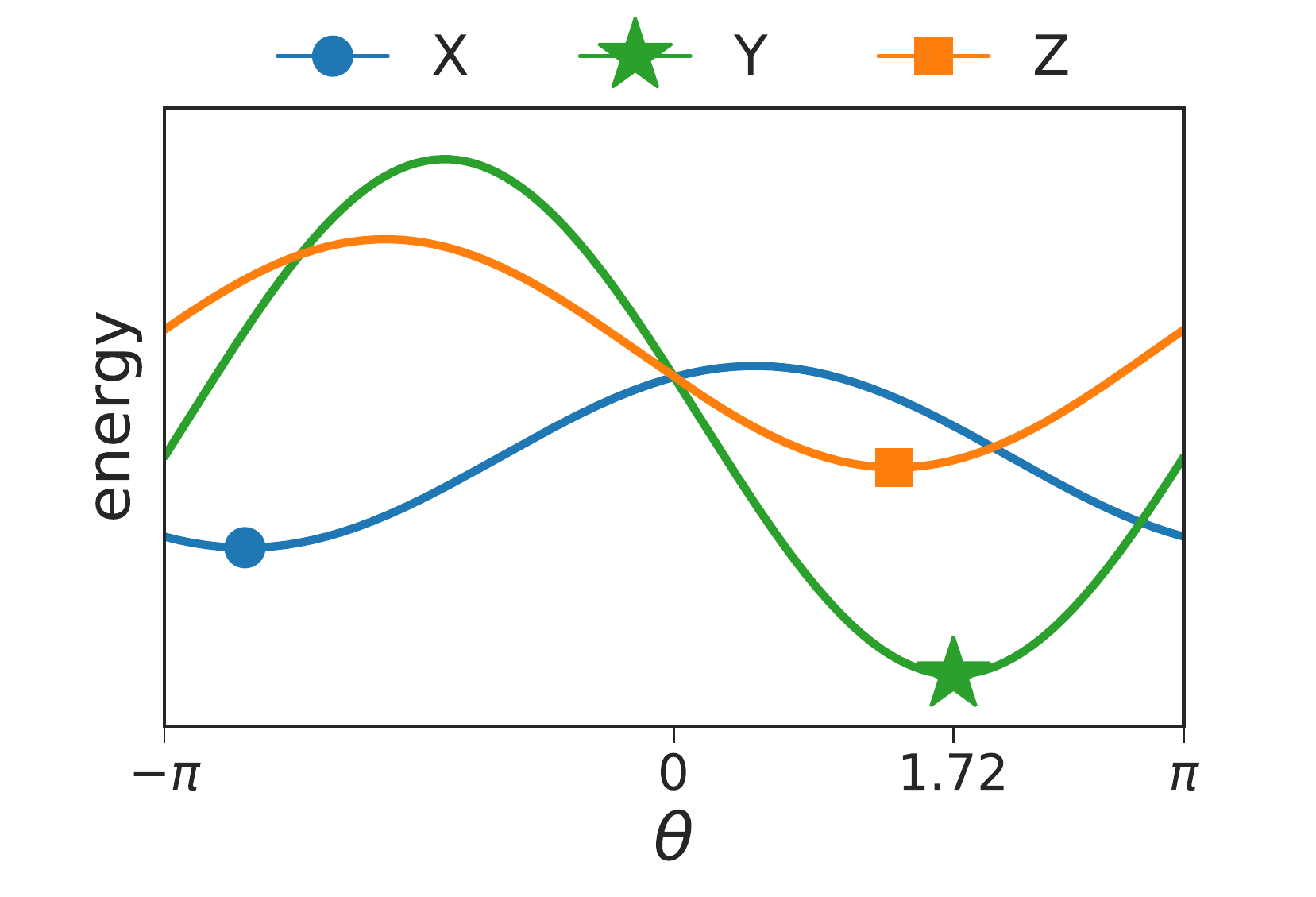}
    \end{minipage}
    \caption{The gradient-free algorithm \texttt{Rotoselect} sequentially adjusts angles of rotation and circuit structure in order to efficiently minimize the energy. For each generator the energy is a sinusoidal function of $\theta$ with period $2\pi$. In this example, the single-qubit gate coloured in yellow has been assigned generator $Y$ and angle of rotation $1.72$. This is the configuration attaining minimal energy, as shown in the right panel.}
    \label{fig:learning} 
  \end{figure}

  The above is very general and indeed suffers from combinatorial explosion due to the $4^n$ possible choices for the generator of each gate. However, in practice we shall still comply with the constraints of the underlying NISQ hardware. Since we are not considering compilation of logical gates to physical ones, only a very small subset of generators is available, namely the native gate set of the hardware. For simplicity, we now consider optimizing the structure of single-qubit gates while employing a fixed layout of two-qubit entangling gates similar to the one in Ref.~\cite{mcclean2018barren}. Figure~\ref{fig:learning} illustrates the circuit layer and shows an example of how the algorithm updates both the generator and the angle. 
  
  When structure optimization is limited to single-qubit gates, the generators can be selected such that ${ H_d \in \{X,Y,Z\} }$. The identity generator is not required because it can be trivially obtained from any other generator by setting the angle to zero. In fact, we can exploit this to reduce the number of circuit evaluations per optimization step. According to Eq.~\eqref{eq:coord_minimum}, each of the three generators requires three circuit evaluations, for a total of $9$ evaluations per optimization step. Recalling that $\phi$ can be chosen at will, we can set $\phi \leftarrow 0$ to obtain the identity gate for all three generators. Since ${ \expval{M}_{0,X} = \expval{M}_{0,Y} = \expval{M}_{0,Z} }$, we can estimate this quantity once, effectively reducing the number of evaluations to $7$ per optimization step. We call this algorithm \texttt{Rotoselect} and summarize it in Algorithm~\ref{alg:rotoselect}. We emphasize that \texttt{Rotosolve} is a classical optimization algorithm for circuit parameters. \texttt{Rotoselect} is an extension to \texttt{Rotosolve} that also allows for some optimization of the structure of the ansatz itself. Neither algorithm is tied to a specific ansatz.
  
  \begin{algorithm}[H]
    \caption{\texttt{Rotosolve}}
    \begin{algorithmic}[1]
      \Require{Hermitian measurement operator $M$ encoding the objective function, parameterized quantum circuit $U$ with fixed structure, stopping criterion}
      \State{Initialize $\theta_d \in (-\pi,\pi]$ for $d=1,\dots,D$ heuristically or at random}
      \Repeat
      \For{$d=1,\dots,D$}
      \State{Select a value $\phi \in \mathbb{R}$ heuristically or at random}
      \State{Fix all angles except the $d$-th one}
      \State{Estimate $\expval{M}_{\phi}$, $\expval{M}_{\phi+\frac{\pi}{2}}$ and $\expval{M}_{\phi-\frac{\pi}{2}}$ from samples}
      \State{$\theta_d \leftarrow \phi - \tfrac{\pi}{2} - \arctantwo(2\expval{M}_{\phi} - \expval{M}_{\phi+\frac{\pi}{2}} - \expval{M}_{\phi-\frac{\pi}{2}} , ~ \expval{M}_{\phi+\frac{\pi}{2}} - \expval{M}_{\phi-\frac{\pi}{2}})$}
      \EndFor
      \Until{stopping criterion is met}
    \end{algorithmic}
    \label{alg:rotosolve}
  \end{algorithm}
  
  \begin{algorithm}[H]
    \caption{\texttt{Rotoselect}}
    \begin{algorithmic}[1]
      \Require {Hermitian measurement operator $M$ encoding the objective function, parameterized quantum circuit $U$, stopping criterion}
      \State Initialize $\theta_d \in (-\pi,\pi]$ and $H_d \in \{ X, Y, Z \}$ for $d=1,\dots,D$ heuristically or at random
      \Repeat
      \For{$d=1,\dots,D$}
      \State{Fix all angles and generators except for the $d$-th gate}
      \For{$P \in \{ X, Y, Z \}$}
      \State{Compute $\theta^*_d(P) = \argmin_{\theta_d} \expval{M}_{\theta_d , P}$ using Eq.~\eqref{eq:coord_minimum} with $\phi \leftarrow 0$}
      \State{Extrapolate $\expval{M}_{\theta_d^*(P), P}$ using the expressions in Appendix~\ref{s:sinusoidal}}
      \EndFor
      \State{$H_d \leftarrow \argmin_P \expval{M}_{\theta_d^*(P), P}$}
      \State{$\theta_d \leftarrow \theta_d^*(H_d)$}
      \EndFor
      \Until{stopping criterion is met}
    \end{algorithmic}
    \label{alg:rotoselect}
  \end{algorithm}
  
  We conclude this section with a note about convergence. On a quantum computer the objective function is stochastic because $ \expval{M}_{\bm{\theta}, \bm{H}} $ is estimated from samples. For implementation in NISQ computers the operator is typically described as the weighted sum of a polynomial number of terms $M = \sum_i w_i M_i$. In this case the expectation is given by ${ \expval{M}_{\bm{\theta}, \bm{H}} = \sum_i w_i \expval{M_i}_{\bm{\theta}, \bm{H}} }$. Assuming an infinite number of measurements the objective function is deterministic and easier to analyze. Bertsekas~\cite{bertsekas1999nonlinear} provides convergence results for coordinate minimization algorithms under rather general conditions and including non-convex objective functions. An analysis along these lines could be attempted for \texttt{Rotosolve}. In \texttt{Rotoselect} the analysis is complicated by the fact that multiple generators could lead to equally good minima. That is, at each given optimization step the solution may not be unique.

  \section{Results}
  
  In this section we first study the impact of structure optimization on the variational quantum eigensolver (VQE) for the Heisenberg model and for Lithium Hydride. \texttt{Rotoselect} differs from \texttt{Rotosolve} only because it employs structure optimization. Any difference in performance between these two algorithms can therefore be attributed to structure optimization. Second, we demonstrate \texttt{Rotoselect} on the IBM Melbourne quantum computer by performing VQE for Hydrogen. Thirdly, we show that an unfavorable initial choice of circuit structure can be boosted using our method. Finally, we compare our algorithms with other state-of-the-art optimizers and find significant improvements in terms of performance and scaling.

  \begin{figure}[H]
    \begin{minipage}[t]{.49\textwidth}
      \centering
      (a)
      \includegraphics[width=\linewidth]{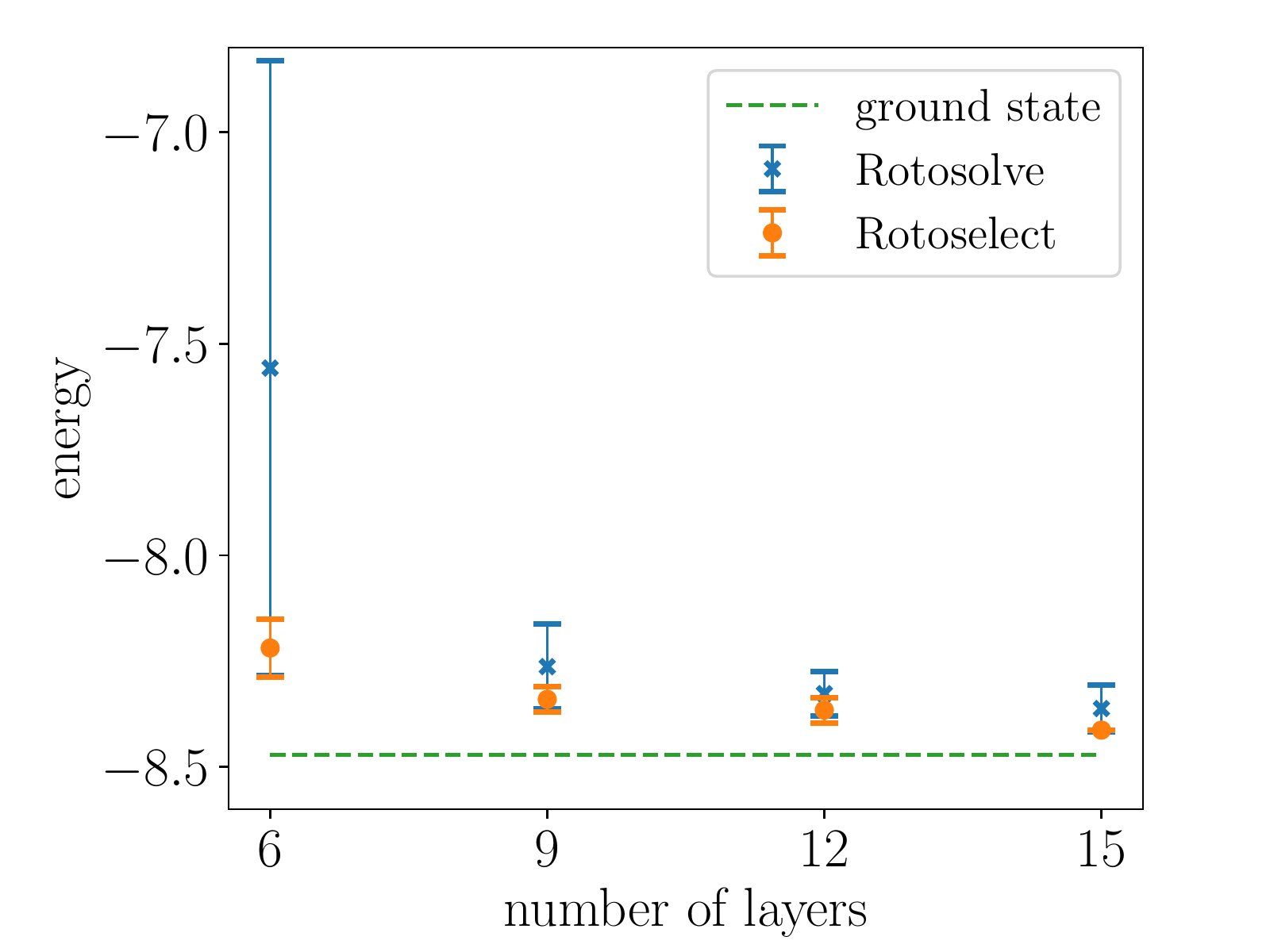}
    \end{minipage}\hfill
    \begin{minipage}[t]{.49\linewidth}
      \centering  
      (b)
      \includegraphics[width=\linewidth]{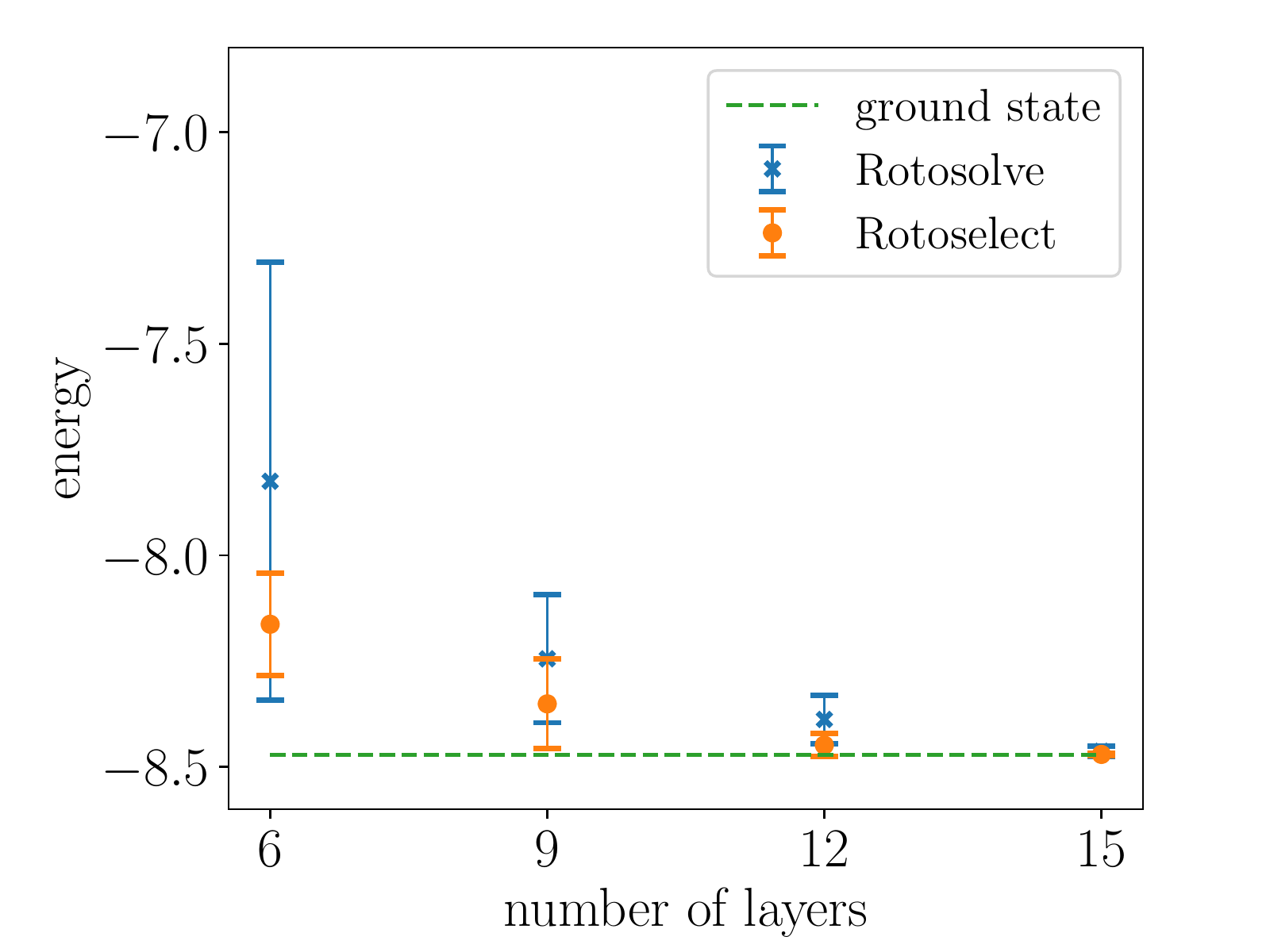}
    \end{minipage}
    \caption{Mean and standard deviation of energy across trials as a function of number of circuit layers comparing \texttt{Rotosolve} and \texttt{Rotoselect} for optimizing a VQE to minimize the energy of the cyclic spin chain Heisenberg model on 5 qubits. In (a) $1000$ measurements for each Hamiltonian term were used to approximate the energy. In (b) the exact energy was calculated. }
    \label{fig:e_v_layers_heisenberg}
  \end{figure}
  
  \begin{figure}[H]
    \begin{minipage}[t]{.49\textwidth}
      \centering
      (a)
      \includegraphics[width=\linewidth]{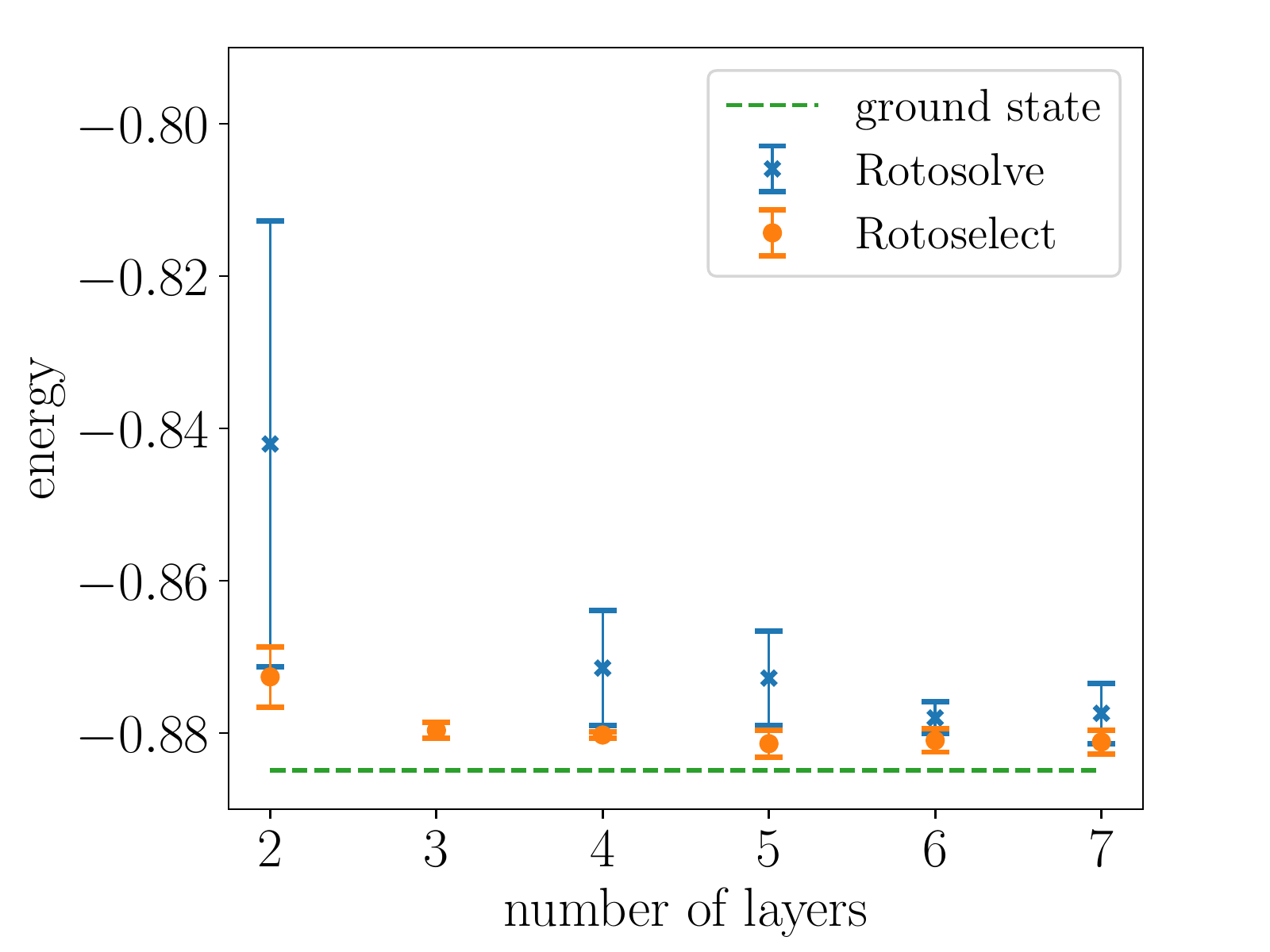}
    \end{minipage}\hfill
    \begin{minipage}[t]{.49\linewidth}
      \centering
      (b)
      \includegraphics[width=\linewidth]{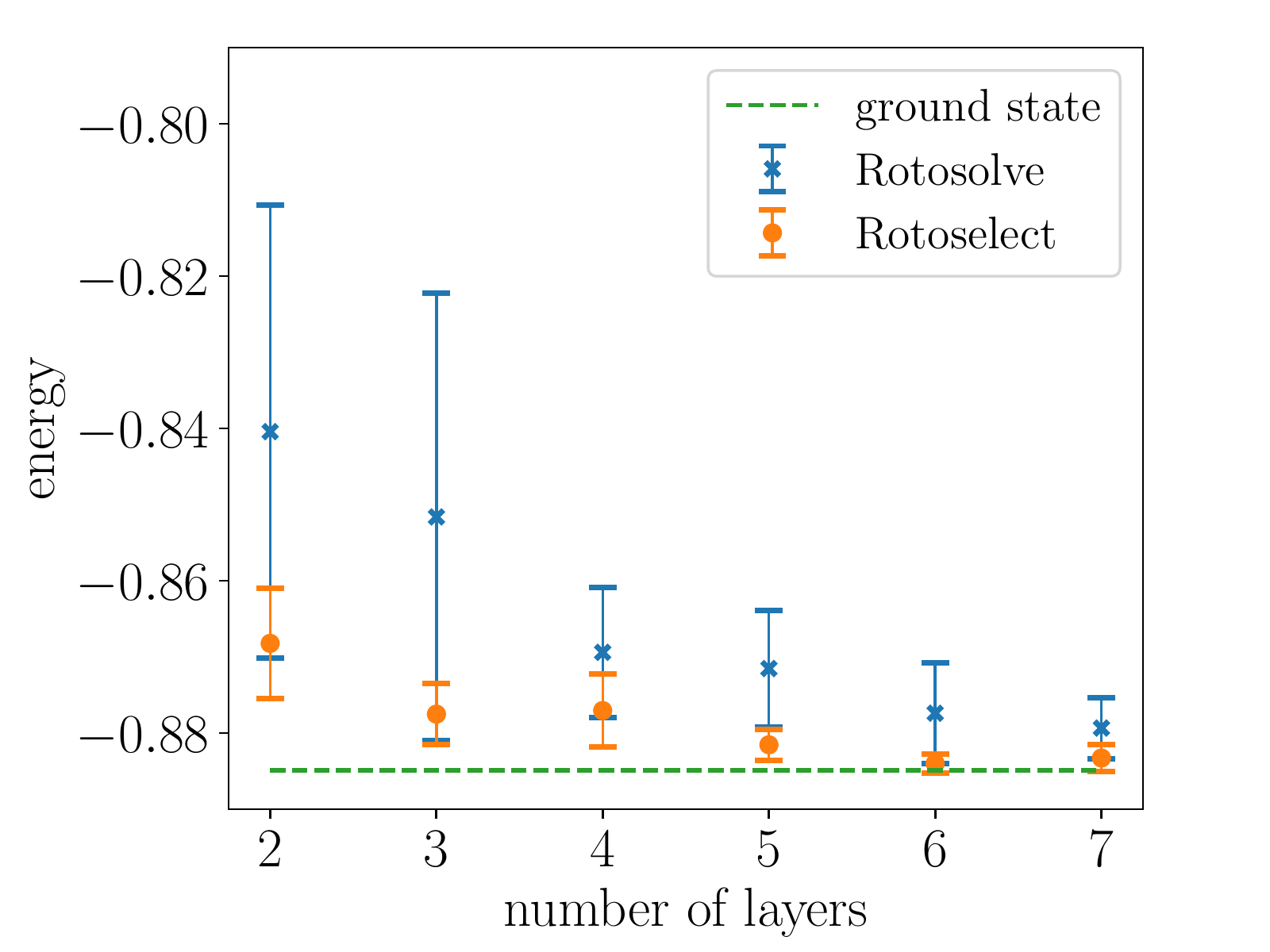}
    \end{minipage}
    \caption{Mean and standard deviation of energy across trials as a function of number of circuit layers comparing \texttt{Rotosolve} and \texttt{Rotoselect} for optimizing a VQE to minimize the molecular Hamiltonian for LiH. In (a) $1000$ measurements for each Hamiltonian term were used to approximate the energy. We do not include the results for \texttt{Rotosolve} on $3$ layers because the mean energy lied above the plotted range due to the presence of an outlier. In (b) the exact energy was calculated.}
    \label{fig:e_v_layers_LiH}
  \end{figure}

  \subsection{Performance on the variational quantum eigensolver}
  
  We considered the problem of finding the ground state energy of the \mbox{$5$-qubit} Heisenberg model on a 1D lattice with periodic boundary conditions and in the presence of an external magnetic field. The corresponding Hamiltonian reads
  \eq{heisenberg}{
    M = J\sum\limits_{(i,j)\in \mathcal{E}}(X_iX_j+Y_iY_j+Z_iZ_j) + h\sum\limits_{i\in \mathcal{V}}Z_i ,
  }
  where $\mathcal{G}=(\mathcal{V},\mathcal{E})$ is the undirected graph of the lattice with $5$ nodes, $J$ is the strength of the spin-spin interactions, and $h$ is to the strength of the magnetic field in the $Z$-direction. For $J/h=1$ the ground state is known to be highly entangled (see Ref.~\cite{kandala2017hardware} for VQE simulations on the Heisenberg model). We chose $J=h=1$.
  
  Circuits were initialized using the strategy described in Ref.~\cite{grant2019initialization}. The circuit form consisted of layers of parameterized single-qubit rotations followed by a ladder of controlled-Z gates~\cite{mcclean2018barren}. An example of this type of layer is shown in Fig.~\ref{fig:learning}. Optimization was performed for $1000$ cycles.
  
  Figure~\ref{fig:e_v_layers_heisenberg} reports mean and standard deviation of the lowest energy encountered during optimization across $10$ trials for circuits with $6,9,12$ and $15$ layers. Panel (a) shows results when $1000$ measurements were used to estimate the expectation of each term in the sum in Eq.~\eqref{eq:heisenberg}, while panel (b) shows oracular results in the limit of infinite measurements. For all number of layers, \texttt{Rotoselect} achieved better mean, standard deviation, and absolute lowest energy than \texttt{Rotosolve}. In particular for the lower depth circuits \texttt{Rotoselect} achieves a much smaller standard deviation than \texttt{Rotosolve}. This is likely due to the fact that the generators are initialized at random and \texttt{Rotosolve} is unable to change an initial bad condition. Given the limited capacity of low-depth circuits, a good choice for the generators appears to be particularly important. As the number of layers increases, both algorithms find better approximations to the ground state. Both algorithms exhibit robustness to the finite number of measurement employed here. 
  
  The experiment was repeated on the $4$-qubit chemical Hamiltonian for Lithium Hydride (LiH) at bond distance. To construct the Hamiltonian we used the coefficients and the Pauli terms given in Ref.~\cite{kandala2017hardware}, Supplementary Material. Figure~\ref{fig:e_v_layers_LiH} shows mean and standard deviation of the lowest energy encountered across $10$ trials for circuits up to $7$ layers. Panel (a) shows results when expectations are estimated from $1000$ measurements, while panel (b) shows results in the limit of infinite measurements. The results are consistent with the previous experiment.

  \subsection{Demonstration on the IBM Melbourne computer}
  
  We compared the performance of \texttt{Rotoselect} on the 14-qubit IBM Melbourne quantum computer against a quantum simulator. For this test we chose to find the ground state energy of the $2$-qubit Hydrogen Hamiltonian, using the coefficients and the Pauli terms given in Ref.~\cite{kandala2017hardware}, Supplementary Material. The circuit had $2$ layers and a total of $4$ adjustable angles. In contrast to previous experiments where we used controlled-$Z$ entangling gates, here we used CNOTs as they belong to the native gate alphabet of this device. 
  
  We also characterized the measurement noise using an off-the-shelf method provided by Qiskit Ignis~\cite{Qiskit}, namely the tensored measurement calibration. A general measurement error calibration prepares each of the $2^n$ basis states and immediately measures them. The statistics are then used to calculate a calibration matrix which is applied in the classical post-processing of subsequent runs. Tensored measurement calibration assumes that errors are local to subsets of qubits, hence reducing the requirements for calculating the calibration matrix. In our experiments, we assumed that measurement errors are local to each qubit. In the following analysis, we do not include the overhead from the calibration since it was performed only once at the beginning of each trial.
  
\begin{figure}[H]
    \centering
    \begin{minipage}[t]{0.5\linewidth}
      \centering
      \includegraphics[width=\linewidth]{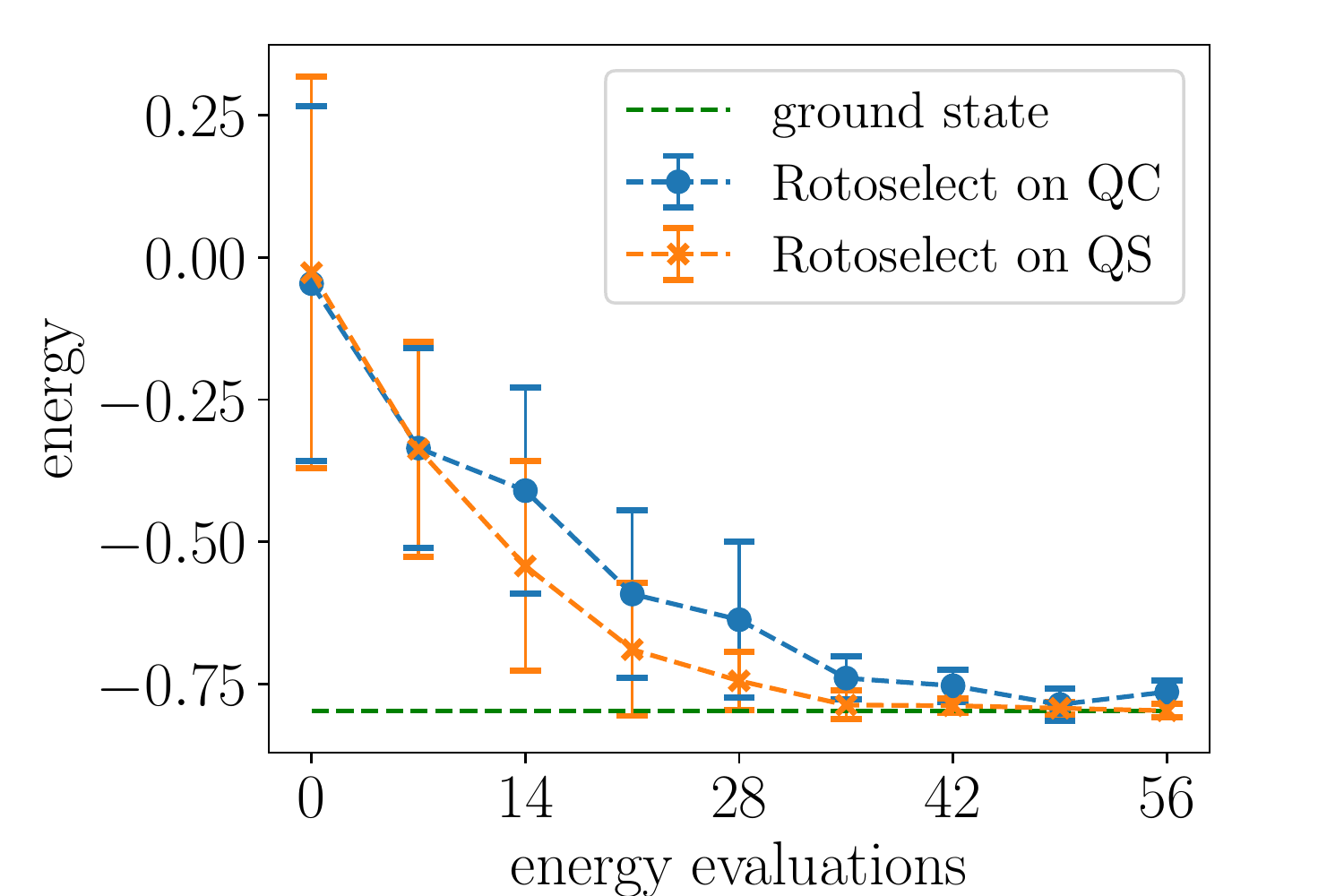}
    \end{minipage}
    \caption{Mean and standard deviation of the energy as a function of the number of evaluations for the Hydrogen Hamiltonian. Blue dots correspond to experiments on the IBM Melbourne quantum computer (QC), and orange crosses corresponds to experiments on the quantum simulator (QS).
    }
    \label{fig:IBMQ_plot}
  \end{figure}

  We ran $5$ trials with randomly initialized circuits and taken $1000$ measurements for each Hamiltonian term when estimating the energy. Figure~\ref{fig:IBMQ_plot} shows the mean and one standard deviation of the energy as a function of the number of evaluations for two full cycles of \texttt{Rotoselect}. Despite error mitigation, results on the quantum device (blue dots) are in average slightly worse than those from the simulator (orange crosses). Yet we were able to get close to the ground state within the $56$ evaluations required to complete two cycles.
  
  \subsection{Optimization of circuits with limited expressibility}
  
  In Ref.~\cite{Aspuru-Guzik2019Expressibility} the authors define ``expressibility'' as the circuit's ability to generate pure states that are well representative of the Hilbert space. To estimate expressibility, they repeatedly sample the adjustable angles of the circuit at random in order to generate a distribution of quantum states; then, they use a suitable divergence to compare this distribution to the uniform distribution of quantum states. Small divergence means high expressibility.
  
  As a function of the number of layers, expressibility improves at different rates depending on the circuit structure. For some choices of structure, additional layers do not yield improvements and expressibility saturates rather quickly. The authors suggest that information about expressibility saturation could be valuable when selecting the circuit structure.
  
\begin{figure}[H]
    \hspace{2ex}
    \begin{minipage}[t]{.45\textwidth}
      \centering
      (a)
      \myCirc{    
        \ket{0}& & \gate{R_Y} & \qw & \qw\oplus   & \qw      & \qw        & \ctrl{1}& \qw& \gate{R_Y}& \qw&\qw &\ctrl{3}& \qw \oplus& \qw&\qw& \\
        \ket{0}& & \gate{R_Y} & \qw & \qw      &  \qw     & \ctrl{1}    &\qw \oplus&
        \qw&  \gate{R_Y}& \qw& \qw& \qw&\ctrl{-1}&\qw \oplus&\qw&\\
        \ket{0}& & \gate{R_Y} & \qw & \qw       & \ctrl{1}  & \qw\oplus &\qw& 
        \qw&  \gate{R_Y} &\qw& \qw \oplus   & \qw&\qw& \ctrl{-1}&\qw \\
        \ket{0}& & \gate{R_Y} & \qw & \ctrl{-3} & \qw\oplus & \qw& \qw&
        \qw&  \gate{R_Y}& \qw &\ctrl{-1}& \qw \oplus & \qw& \qw&\qw&
      }
    \end{minipage}
    \hspace{8ex}
    \begin{minipage}[t]{.45\textwidth}
      \centering
      (b)
      \includegraphics[width=\linewidth]{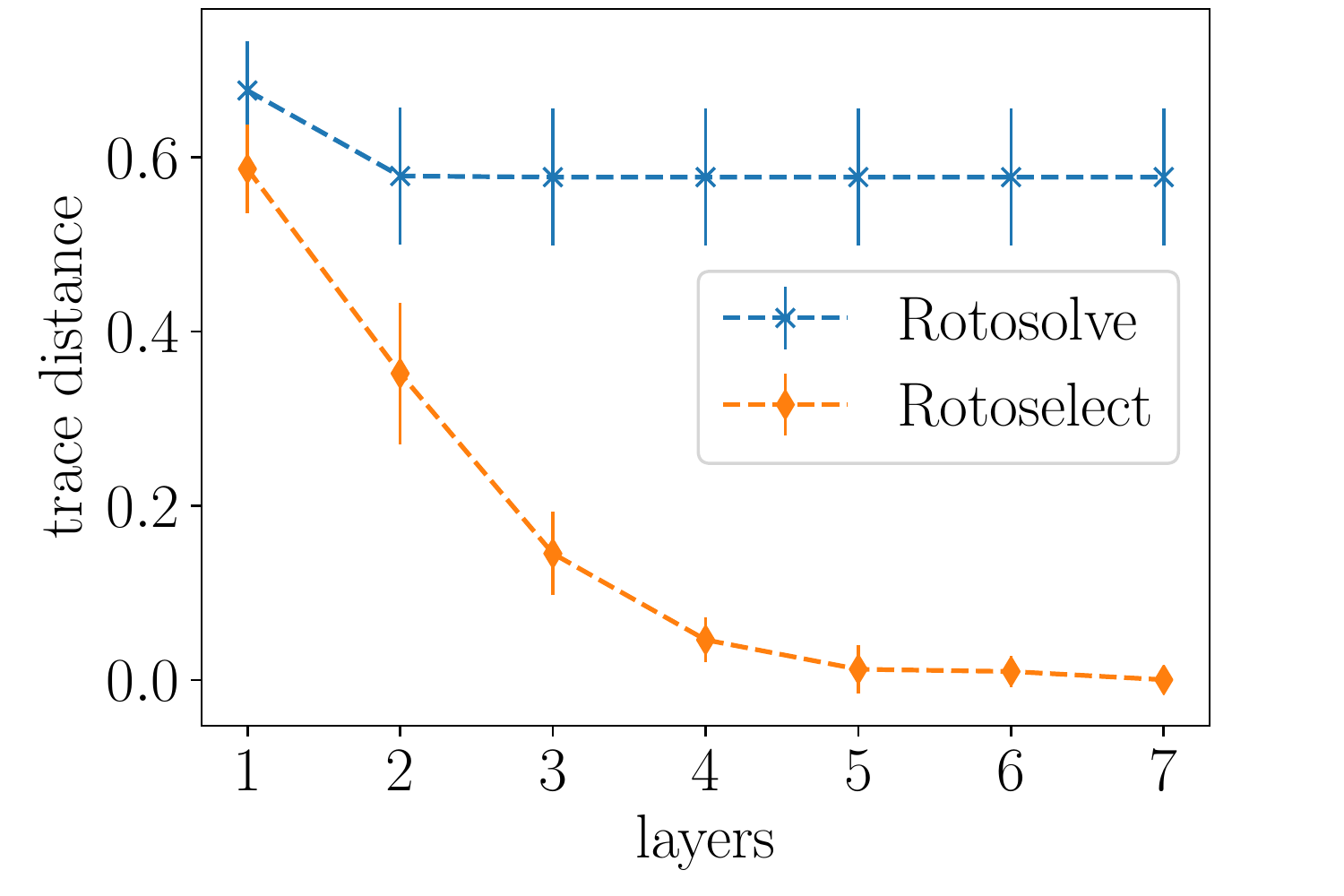}
    \end{minipage}
    \caption{(a) Diagram for circuit $\#15$ from Ref.~\cite{Aspuru-Guzik2019Expressibility}. (b) Mean and standard deviation of the trace distance to uniformly random states as a function of number of circuit layers for \texttt{Rotosolve} and \texttt{Rotoselect} on \mbox{circuit $\#15$}.}
    \label{fig:express_circs}
  \end{figure}

  In this section, we use expressibility to select an unfavorable circuit and show that structure optimization can turn it into a useful one. In particular, we required a circuit structure for which expressibility is poor and saturates quickly. We chose circuit $\#15$ from the pool of circuits studied in Ref.~\cite{Aspuru-Guzik2019Expressibility}. Figure~\ref{fig:express_circs}~(a) shows the corresponding circuit diagram for $4$ qubits. Our task consisted of maximizing the overlap of the state generated by the circuit with a target state sampled uniformly at random. To avoid potential misunderstanding, we stress that our task is not that of maximizing the expressibility, although the two may be related. In practice, we minimized the energy for operator $M = -\dyad{\phi}$ where $\ket{\phi}$ is the random target state.
  
  For this simulation, we used the exact energy and we varied the number of layers in $\{1,\ldots,7\}$. We stopped the algorithms after 50 cycles. Figure~\ref{fig:express_circs}~(b) shows the average trace distance and standard deviation as a function of layers across $10$ random target states for circuit $\#15$. \texttt{Rotosolve} is not able to take the trace distance below a certain value, even when adding more layers. On the other hand, \texttt{Rotoselect} achieves lower trace distance for all choices of the number of layers and approaches zero for $7$ layers. This indicates that an unfavorable initial choice of circuit structure can be boosted using our method.
  
  \subsection{Comparison of optimization algorithms and scaling}
  
  In this experiment we compare the optimization time and optimization time scaling in the system size for \texttt{Rotosolve}, \texttt{Rotoselect}, \texttt{SPSA}~\cite{spall1992multivariate} and gradient descent with \texttt{Adam}~\cite{kingma2014adam}.
  
\begin{figure}[H]
    \begin{minipage}[t]{.49\textwidth}
      \centering
      (a)
      \includegraphics[width=\linewidth]{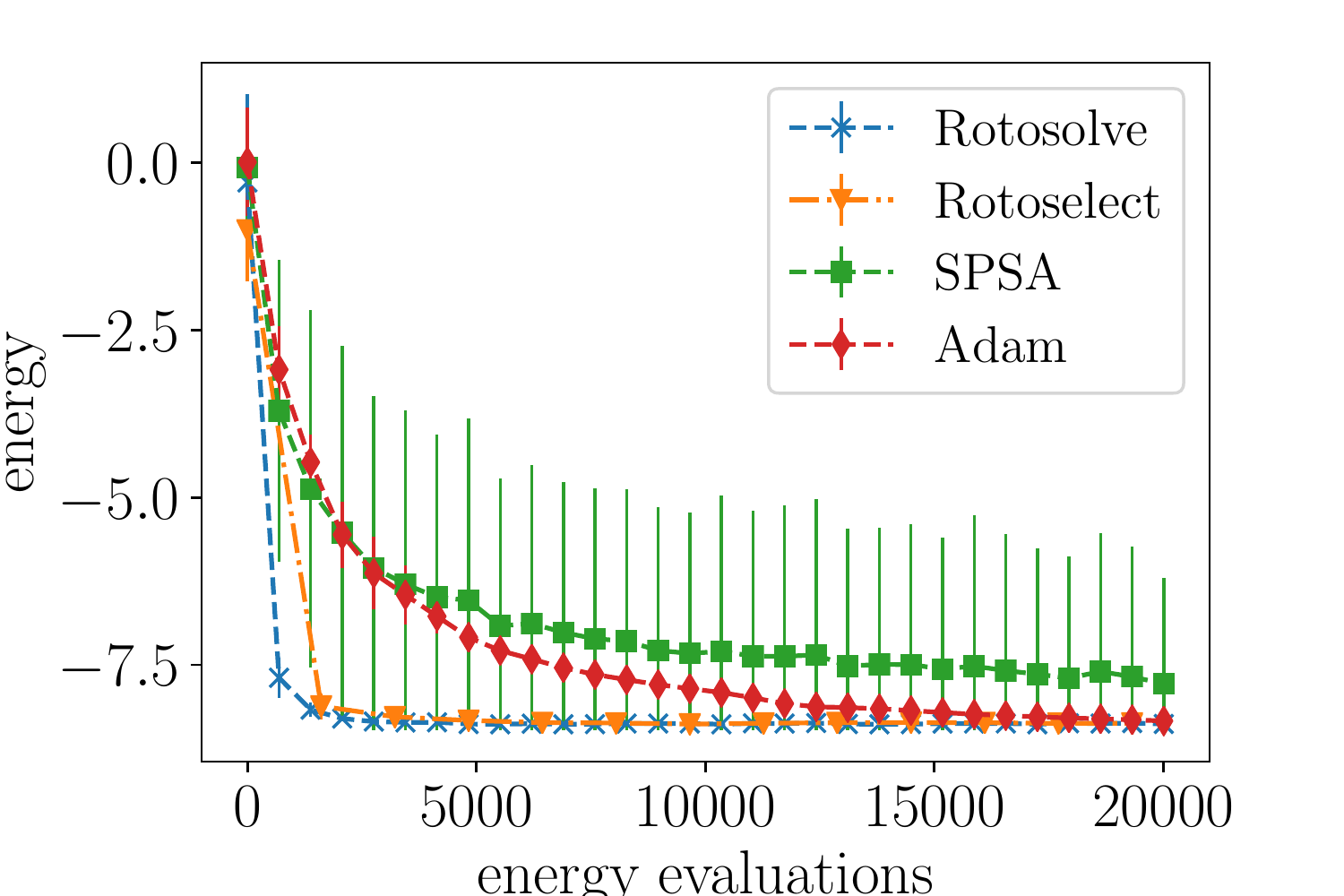}
    \end{minipage}\hfill
    \begin{minipage}[t]{0.49\linewidth}
      \centering
      (b)
      \includegraphics[width=\linewidth]{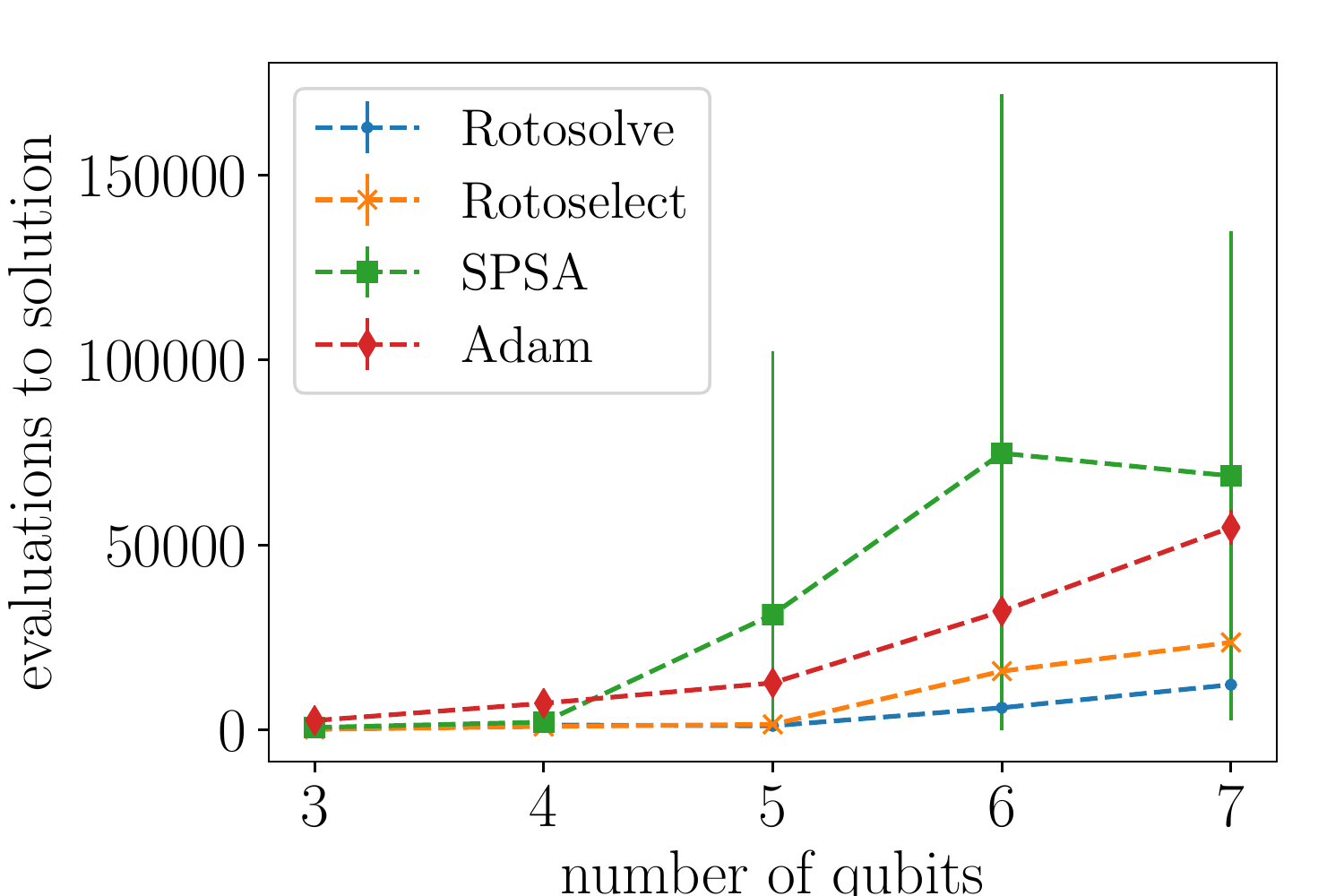}
    \end{minipage}
    \caption{A comparison of \texttt{Rotosolve}, \texttt{Rotoselect}, \texttt{SPSA} and \texttt{Adam}. (a) Energy as a function of number of energy evaluations for the $5$-qubit Heisenberg cyclic spin chain. (b) Number of energy evaluations to solution as a function of the number of qubits for the Heisenberg cyclic spin chain.}
    \label{fig:comparison}
  \end{figure}

  Figure~\ref{fig:comparison}~(a) shows the energy and standard deviation as a function of the number of energy evaluations across $5$ trials for the $5$-qubit Heisenberg cyclic spin chain described in the Results section. The depth of the circuit was $30$ layers. The exact energy was used to perform updates. The learning rate for \texttt{Adam} was set to $0.05$. The hyperparameter for \texttt{SPSA} were the same as in Ref.~\cite{kandala2017hardware}. Both \texttt{Rotosolve} and \texttt{Rotoselect} converged with significantly fewer energy evaluations than \texttt{Adam} or \texttt{SPSA}.
  
  Figure~\ref{fig:comparison}~(b) shows the mean and standard deviation for the number of evaluations needed to find the ground state of the Heisenberg cyclic spin chain as a function of the size of the system. The ground state was considered found if the candidate solution had energy within $2\%$ from the actual ground state energy. More precisely, this threshold is in relation to the normalized distance between the candidate energy and the minimum eigenvalue expressed by  $\frac{\expval{M} -E_{\min}}{E_{\max}-E_{\min}}$.
  It should be noted, that in the experiments there was an additional stop condition, which stopped algorithm after $100,000$ cycles. Five trials were performed to estimate each mean and standard deviation. $1000$ measurements were performed for each Hamiltonian term in order to approximate the energy for each evaluation. Circuit depth was ${3n^2}/{2}+2n$ for an even number of qubits and ${3(n^2-1)}/{2} +2n$ for odd where $n$ was the number of qubits.
  
  Both \texttt{Rotosolve} and \texttt{Rotoselect} converged faster than \texttt{Adam} or \texttt{SPSA} for the number of qubits tested, indicating a favorable scaling.
  
  \section{Discussion}
  
  The proposed algorithms can be extended in a number of ways. In the context of circuit optimization, \texttt{Rotosolve} can be generalized to find minimizers of $K$ angles of rotation at the same time and at the cost of evaluating $3^K$ circuits. While this is an exponential cost, for small $K$ this approach is computationally feasible and may provide an advantage. In Ref.~\cite{parrish2019jacobi} the authors explore this idea for $K \in \{1,2,3\}$ where subsets of angles are chosen at random. This approach belongs to the class of algorithms called coordinate block minimization~\cite{wright2015coordinate}.
  
  In the context of circuit structure optimization, \texttt{Rotoselect} can be generalized to incrementally grow circuits from scratch rather than starting from random initial structures. This is similar in spirit to \mbox{Adapt-VQE~\cite{grimsley2018adapt}}, but with the advantage that each new gate is optimized efficiently in closed form rather than using gradient-based optimization. Furthermore, we could efficiently remove gates by assessing whether they are contributing to the solution (e.g., a redundant gate would generate sinusoidal forms that appear to be flat).
  
  Another interesting extension is to use \texttt{Rotoselect} to learn the optimal connectivity layout for the entangling gates. As an example consider trapped ion computers that can implement fully connected layers of M{\o}lmer-S{\o}rensen gates~\cite{sorensen1999quantum}. This choice provides high expressive power in low-depth circuits~\cite{benedetti2019generative}, but must be balanced against the potentially slow clock speed of these entangling gates~\cite{linke2017experimental} and potential gate errors. Our algorithm could find this sweet spot. Since with $n$ qubits we must evaluate $n(n-1)/2$ choices for non-directional two-qubit gates in a layer, this approach is also efficient. We discuss other generalizations in Appendix~\ref{s:generalizations}.
  
  Heuristics can be used to speedup our methods. The most striking example consists of reusing information to reduce the number of energy evaluations. By appropriately choosing the offset $\phi$ in Eq.~\eqref{eq:coord_minimum}, one obtains an angle for which the energy is already known from the previous update. Thus one only needs to evaluate the remaining two energies to perform the current update. By applying this trick systematically one reduces the number of evaluations from 3 to 2 for \texttt{Rotosolve}, and from 7 to 6 for \texttt{Rotoselect}. While the result is exact in the limit of infinite number of measurement, in realistic scenarios this approximation introduces noise and its effects should be investigated in future work.
  
  In our simulations we also found that \texttt{Rotoselect} tends to change the structure only in the early stage of optimization. One could detect this and switch from \texttt{Rotoselect} to \texttt{Rotosolve} further reducing the number of circuit evaluations.
  
  Quantum circuit structure optimization provides an efficient means for improving the expressivity of low-depth quantum circuits with only a small computational overhead. This characteristic makes the described methods particularly suitable for deployment on near-term quantum computers where circuit depth and optimization time are significant bottlenecks.
  
  \section*{Acknowledgements}
  M.B. is supported by the UK Engineering and Physical Sciences Research Council (EPSRC) and by Cambridge Quantum Computing Limited (CQC). E.G. is supported by ESPRC [EP/P510270/1]. M.O. acknowledges support from Polish National Science Center scholarship 2018/28/T/ST6/00429. We thank Leonard Wossnig for helpful discussions. We gratefully acknowledge the support of NVIDIA Corporation with the donation of the Titan Xp GPU used for this research. This research was supported in part by PLGrid Infrastructure.

  \printbibliography
  
  \appendix
  
  \section{Sinusoidal form of expectation values}
  \label{s:sinusoidal}
  
  Recall that we consider circuits $U = U_D \cdots U_1$ where each gate is either fixed, e.g., the controlled-Z, or parameterized. We consider parameterized gates of the kind $U_d = \exp(-i \tfrac{\theta_d}{2} H_d)$, where $\theta_d \in (-\pi, \pi]$ and the generator $H_d$ is a Hermitian and unitary matrix such that $H_d^2=I$. Using the definition of matrix exponential we obtain
  \eq{gate}{
    U_d &= \sum_{k=0}^\infty \frac{ (-i)^k (\tfrac{\theta_d}{2})^k H_d^k }{k!} \\
    &= \sum_{k=0}^\infty \frac{(-i)^{2k} (\tfrac{\theta_d}{2})^{2k} H_d^{2k}}{(2k)!} + \sum_{k=0}^\infty \frac{ (-i)^{2k+1} (\tfrac{\theta_d}{2})^{2k+1} H_d^{2k+1}}{(2k+1)!} \\
    &= \sum_{k=0}^\infty \frac{(-1)^{k} (\tfrac{\theta_d}{2})^{2k}}{(2k)!} I - i \sum_{k=0}^\infty \frac{(-1)^{k} (\tfrac{\theta_d}{2})^{2k+1} }{(2k+1)!} H_d \\
    &= \cos ( \tfrac{\theta_d}{2} ) I - i\sin (\tfrac{\theta_d}{2} ) H_d .
  }
  As an example of generators with this property, we could use tensor products of Pauli matrices ${H_d \in \{ I, X, Y, Z\}^{\otimes n}}$, where $n$ is the number of qubits.
  
  We apply the circuit to a fiducial state $\bar{\rho}$ and then measure a Hermitian operator $\bar{M}$ encoding the objective function. From measurement outputs we can estimate the expectation
  \eq{exp_full}{
    \expval{\bar{M}} = \tr(\bar{M} U_D \cdots U_d \cdots U_1 \bar{\rho} U_1^\dag \cdots U_d^\dag \cdots U_D^\dag) .
  }
  We want to analyze this expectation as a function of a single parameter $\theta_d$. To simplify the notation we absorb all gates before $U_d$ in the density operator, that is $\rho =  U_{d-1} \cdots U_1 \bar{\rho} U_1^\dag \cdots U_{d-1}^\dag$. Similarly, we absorb all gates after $U_d$ in the measurement operator $M = U_{d+1}^\dag \cdots U_D^\dag \bar{M} U_D \cdots U_{d+1}$. This can be done because unitary transformations preserve the Hermiticity of both density and measurement operators. Using this notation Eq.~\eqref{eq:exp_full} can be written as $\expval{M} = \tr(M U_d \rho U_d^\dag)$. 
  
  In the following discussion we will need to evaluate the expectation at different parameter values for gate $U_d$. A further change in notation helps. From now on, we drop index $d$ and use subscripts to indicate parameter value. Using the new notation we write
  \eq{exp_1}{
    \expval{M}_\theta &= \tr \left( M U_\theta \rho U_\theta^\dag \right) \\
    &= \tr \left( M \left( \cos \left( \tfrac{\theta}{2} \right) I - i \sin \left( \tfrac{\theta}{2} \right) H \right) \rho \left( \cos \left( \tfrac{\theta}{2} \right) I + i \sin \left( \tfrac{\theta}{2} \right) H \right) \right) \\
    &= \cos^2 \left( \tfrac{\theta}{2} \right) \tr \left( M\rho \right) + i \sin \left( \tfrac{\theta}{2} \right) \cos \left( \tfrac{\theta}{2} \right) \tr \left( M \left[ \rho, H \right] \right) +\sin^2 \left( \tfrac{\theta}{2} \right) \tr \left( M H \rho H \right) .
  }
  
  Let us inspect the third line of this equation. First, we note that $\tr(M \rho)$ is simply the expectation when the circuit is evaluated at $\theta=0$. That is, $\expval{M}_0 = \tr(M \rho)$. 
  Second, the term $i\tr(M [\rho, H])$ can be written as the difference of two expectations. Indeed, using two independent circuits $U_{\pm\frac{\pi}{2}} = \exp(\mp i \tfrac{\pi}{4}H) = \tfrac{1}{\sqrt{2}} (I \mp iH)$ we have
  \eq{commutator}{
    \expval{M}_{\frac{\pi}{2}} - \expval{M}_{-\frac{\pi}{2}} &= \tr\left(M U_{\frac{\pi}{2}} \rho U_{\frac{\pi}{2}}^\dag\right) - \tr\left(M U_{-\frac{\pi}{2}} \rho U_{-\frac{\pi}{2}}^\dag\right) \\
    &= \tfrac{1}{2} \left( \tr \left( M \left(I - iH\right) \rho \left(I + iH\right) \right) - \tr \left( M \left(I + iH\right) \rho \left(I - iH\right) \right) \right)  \\
    &= -i\tr\left(M H \rho\right) + i\tr\left(M \rho H\right) \\
    &= i\tr\left(M [\rho, H]\right) .
  }
  
  Third, the last term $\tr \left( M H \rho H \right)$ is the expectation obtained evaluating the circuit at $\theta=\pi$. That is, using circuit $U_\pi = \exp(-i\tfrac{\pi}{2}H) = -iH$ we have $\expval{M}_{\pi} = \tr \left( M H \rho H \right)$.
  
  Putting these three pieces back in Eq.~\eqref{eq:exp_full} and using well known trigonometric identities, we obtain
  \eq{exp_2}{
    \expval{M} \def \expval{M}_\theta &= \cos^2\left(\tfrac{\theta}{2}\right) \expval{M}_0 + \sin\left(\tfrac{\theta}{2}\right)\cos\left(\tfrac{\theta}{2}\right) \left( \expval{M}_{\frac{\pi}{2}} - \expval{M}_{-\frac{\pi}{2}} \right) +\sin^2\left(\tfrac{\theta}{2}\right) \expval{M}_\pi \\
    &= \tfrac{1+\cos\left(\theta\right)}{2} \expval{M}_0 + \tfrac{\sin\left(\theta\right)}{2} \left( \expval{M}_{\frac{\pi}{2}} - \expval{M}_{-\frac{\pi}{2}} \right) + \tfrac{1-\cos\left(\theta\right)}{2} \expval{M}_\pi\\
    &= \tfrac{\cos\left(\theta\right)}{2} \left( \expval{M}_0 - \expval{M}_\pi \right) + \tfrac{ \sin\left(\theta\right) }{2} \left( \expval{M}_{\frac{\pi}{2}} - \expval{M}_{-\frac{\pi}{2}}\right) + \tfrac{1}{2} \left( \expval{M}_0 + \expval{M}_\pi \right) .
  }
  
  We now use the identity $a \cos(x) + b \sin(x) = \sqrt{a^2 + b^2} \sin(x + \arctan(\frac{a}{b}))$ to obtain a compact expression. The expectation value then reads
  \eq{exp_3}{ 
    \expval{M}_\theta &= A \sin(\theta + B) +C , \\
    A &= \tfrac{1}{2} \sqrt{ \big( \expval{M}_0 - \expval{M}_\pi \big)^2 + \big( \expval{M}_{\frac{\pi}{2}} -\expval{M}_{-\frac{\pi}{2}} \big)^2 }, \\
    B &= \arctantwo \left( \expval{M}_0 - \expval{M}_\pi , \expval{M}_{\frac{\pi}{2}} - \expval{M}_{-\frac{\pi}{2}} \right ), \\
    C &= \tfrac{1}{2} \left( \expval{M}_0 + \expval{M}_\pi \right) .
  }
  
\begin{figure}
    \begin{minipage}[t]{.46\textwidth}
      \centering 
      (a)
      \includegraphics[width=\linewidth]{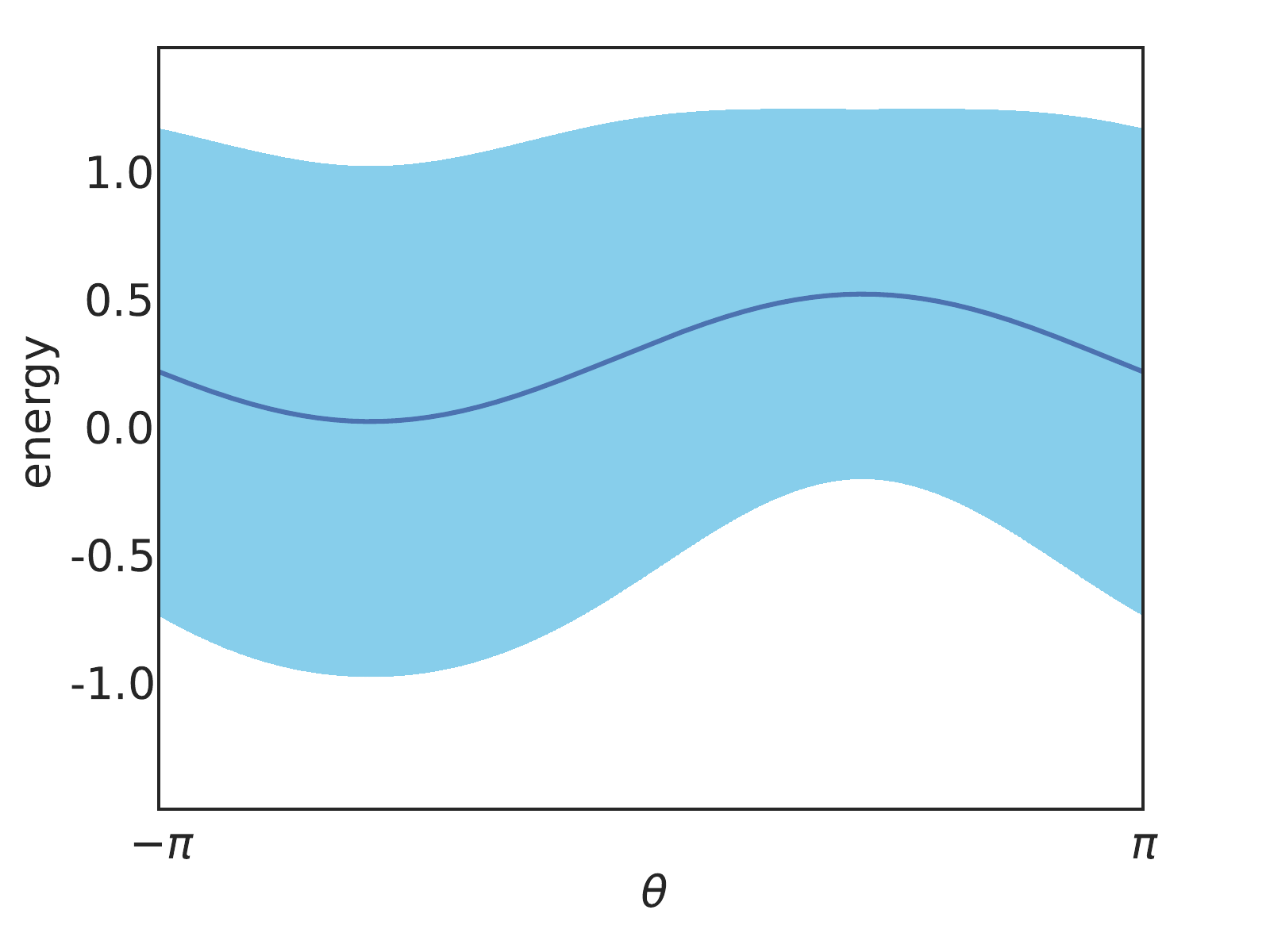}
    \end{minipage}\hfill
    \begin{minipage}[t]{0.46\linewidth}
      \centering 
      (b)
      \includegraphics[width=\linewidth]{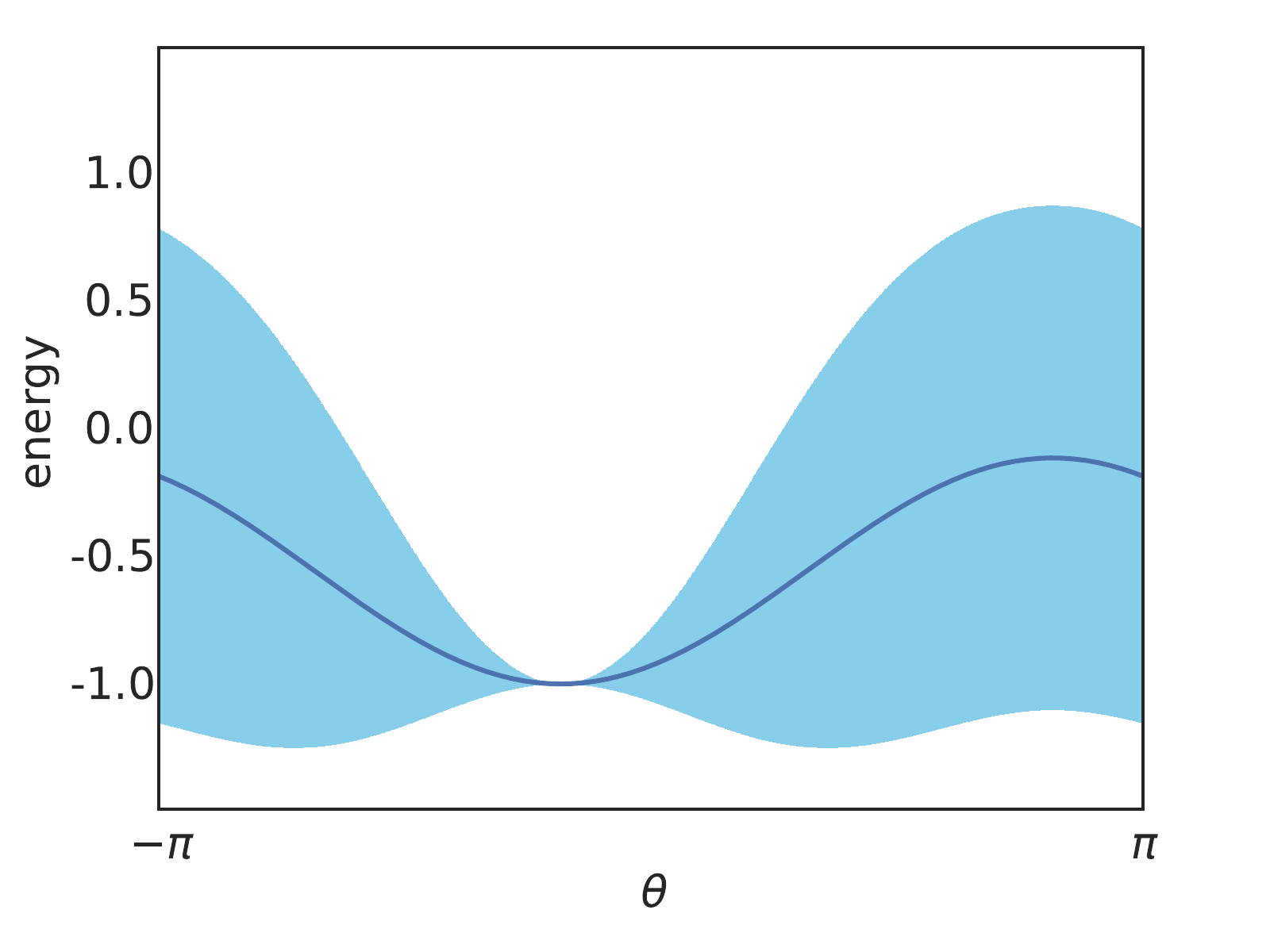}
    \end{minipage}
    \caption{Expectation and variance of Hermitian observable $ZXZX$ as a function of two different angles of rotation in a random circuit. The expectation value is always of sinusoidal form with period $2\pi$. (a) The variance is large because this angle cannot yield an eigenstate of $ZXZX$. (b) Both expectation and variance are minimized by setting this second angle to $\theta^* \approx -0.5$.}
    \label{fig:sinusoidal_form}
  \end{figure}

  That is, the expectation of $M$ as a function of a single parameter has sinusoidal form with amplitude $A$, phase $B$, intercept $C$, and period $2\pi$. An example is shown in Figure~\ref{fig:sinusoidal_form}.
  
  Note that we must use the $\arctantwo$ function in order to correctly handle the sign of numerator and denominator, as well as the case where the denominator is zero. To see why, assume $H$ commutes either with $M$ or $\rho$. Then, Eq.~\eqref{eq:commutator} evaluates to zero and triggers a division by zero in the standard $\arctan$. 
  
  Now, from the graph of the sine function, it is easy to locate the minima at ${ \theta^* = - \frac{\pi}{2} - B + 2\pi k}$ for all $k \in \mathbb{Z}$. The estimator for $B$ in Eq.~\eqref{eq:exp_3} seems to require four distinct circuit evaluations. Using simple trigonometry we generalize the estimator and show that only three evaluations are required. Let us write
  \eq{trig_trick}{
    \expval{M}_{\phi+\tfrac{\pi}{2}} &= A \sin(\phi + \tfrac{\pi}{2} + B) +C \\
    &= -A \sin(\phi -\tfrac{\pi}{2} + B) +C \\
    &= -\expval{M}_{\phi-\tfrac{\pi}{2}} +2C .
  }
  From this we obtain the general estimator for $C$
  \eq{general_C}{
    C = \tfrac{1}{2} \left( \expval{M}_{\phi+\tfrac{\pi}{2}}+\expval{M}_{\phi - \tfrac{\pi}{2}} \right) .
  }
  We can write
  \eq{trig_trick_3}{
    \tan\left(\phi + B\right) &= \frac{\sin\left(\phi + B\right)}{\sin\left(\phi + B - \frac{\pi}{2}\right)} = \frac{\expval{M}_\phi - C }{\expval{M}_{\phi - \frac{\pi}{2}} - C} = 
    \frac{2\expval{M}_\phi - \expval{M}_{\phi+\frac{\pi}{2}} - \expval{M}_{\phi- \frac{\pi}{2}}}{\expval{M}_{\phi + \frac{\pi}{2}} - \expval{M}_{\phi - \frac{\pi}{2}}} .
  }
  Taking the inverse tangent we obtain the general estimator for $B$ requiring only three circuit evaluations
  \eq{general_B}{
    B &= \arctantwo \Big( 2\expval{M}_{\phi} - \expval{M}_{\phi+\frac{\pi}{2}} - \expval{M}_{\phi-\frac{\pi}{2}} , ~ \expval{M}_{\phi +\frac{\pi}{2}} - \expval{M}_{\phi-\frac{\pi}{2}} \Big) -\phi .
  }
  Finally, we express the minimizer in closed form 
  \eq{final_form}{
    \theta^* &= \argmin_\theta \expval{M}_\theta \\
    &= - \tfrac{\pi}{2} - B + 2\pi k\\
    &= \phi - \tfrac{\pi}{2} - \arctantwo \Big( 2\expval{M}_{\phi} - \expval{M}_{\phi+\frac{\pi}{2}} - \expval{M}_{\phi-\frac{\pi}{2}} , ~ \expval{M}_{\phi +\frac{\pi}{2}} - \expval{M}_{\phi-\frac{\pi}{2}} \Big) +2\pi k ,
  }
  where $\phi \in \mathbb{R}$ and $k \in \mathbb{Z}$. Note that in practice we chose $k$ such that $\theta^* \in (-\pi,\pi]$.\\
  At this point we may also want a general estimator for $A$. It is easy to verify that
  \eq{general_A}{
    A = \tfrac{1}{2} \sqrt{ \big( 2\expval{M}_{\phi} - \expval{M}_{\phi+\frac{\pi}{2}} - \expval{M}_{\phi-\frac{\pi}{2}} \big)^2 + \big( \expval{M}_{\phi+\frac{\pi}{2}} -\expval{M}_{\phi-\frac{\pi}{2}} \big)^2 } .
  }
  This is useful when we need to extrapolate the energy attained by the minimizer, $\expval{M}_{\theta^*} = -A +C$, and can be done at no additional cost.
  In summary, Eqs.~\eqref{eq:general_C},\eqref{eq:general_B} and~\eqref{eq:general_A} completely characterize the sinusoidal form of the energy and allow us to estimate both minimizer and minimum in closed form.

  \section{A few generalizations}
  \label{s:generalizations}
  
  We now discuss some generators $H_d$ for which \texttt{Rotosolve} applies. Our derivation above is rather general, but in practice Algorithms~\ref{alg:rotosolve} and~\ref{alg:rotoselect} rely on the canonical axes of rotation, i.e, tensor products of Pauli matrices.
  
  Our first generalization applies to single-qubits gates. Recall that in order to obtain a sinusoidal form of expectation values we need $H_d^2=I$. For single-qubit gates this condition is met by any ${H_d = c_x X + c_y Y + c_z Z}$ such that $(c_x, c_y, c_z) \in \mathbb{R}^3$ is a unit vector. To see why,
  \eq{gen1}{
    H_d^2 &= (c_x X + c_y Y + c_z Z)(c_x X + c_y Y + c_z Z) \\
    &= c_x^2 X^2 + c_x c_y XY + c_x c_z XZ + c_y c_x YX + c_y^2 Y^2 + c_y c_z YZ + c_z c_x ZX + c_z c_y ZY + c_z^2 Z^2\\
    &= c_x^2 I + i c_x c_y Z - i c_x c_z Y - i c_y c_x Z + c_y^2 I + i c_y c_z X + i c_z c_x Y -i c_z c_y X + c_z^2 I \\
    &= (c_x^2 + c_y^2 + c_z^2) I = I ,}
  where in the third line we used well-known identities for the Pauli matrices. Therefore, $\expval{M}_{\theta_d}$ has sinusoidal form for single-qubit gates of the kind ${U_d = \exp(-i\frac{\theta_d}{2} (c_x X + c_y Y + c_z Z) )}$.
  
  This suggests a version of \texttt{Rotosolve} where a unit vector of coefficients is sampled at random, and where $\theta_d$ is found using Algorithm~\ref{alg:rotosolve}. This version has the potential of finding interesting axes of rotation. Note that a NISQ implementation may require compilation of $U_d$ to lower-level gates.
  
  Our second generalization applies to gates acting on an arbitrary number of qubits. Starting from tensor products of $n$ Pauli matrices, i.e.,  $H_d \in \{I, X, Y, Z\}^{\otimes n}$, any unitary transformation $V$ provides an alternative valid generator $H_d^\prime = V H_d V^\dag$. To see why,
  \eq{gen2}{
    (H_d^\prime)^2 &= V H_d V^\dag V H_d V^\dag = V H_d^2 V^\dag = VV^\dag = I.
  }
  The result is an $n$-qubit gate of the form $U_d = \exp(-i\tfrac{\theta_d}{2} V H_d V^\dag )$, where $\theta_d$ is again found using Algorithm~\ref{alg:rotosolve}. We leave the question of how to exploit gates of this kind for future work.

\end{document}